\documentclass[12pt]{article}
\begin{document}
\title{The effect of annealing on the elastoplastic response 
of isotactic polypropylene}

\author{Aleksey D. Drozdov\footnote{Corresponding author, fax: +45 9815 3030, 
E--mail: drozdov@iprod.auc.dk} and Jesper deC. Christiansen\\
Department of Production\\ 
Aalborg University\\
Fibigerstraede 16, DK--9220 Aalborg, Denmark}
\date{}
\maketitle

\begin{abstract}
Four series of tensile loading--unloading tests are performed 
on isotactic polypro\-py\-lene in the sub-yield domain of deformations
at room temperature.
In the first series, injection-molded specimens are used as
produced, whereas in the other series the samples are annealed
for 24 h at 120, 140 and 160 $^{\circ}$C,
which covers the low-temperature region and an initial part of the 
high-temperature region of annealing temperatures.
A constitutive model is developed for the elastoplastic behavior
of a semicrystalline polymer.
The stress--strain relations are determined by five adjustable
parameters that are found by fitting the experimental data.
The effect of annealing is analyzed on the material constants.
\end{abstract}
\vspace*{10 mm}

\newpage

\section{Introduction}

This study deals with the elastoplastic behavior of injection-molded 
isotactic polypropylene (iPP) in isothermal uniaxial tests 
with small strains.
Plastic deformations and yielding of iPP have been the focus of attention 
in the past decade, see, e.g., 
\cite{AGU95,UU97,CCG98,NT99,SSE99a,SSE99b,NT00,LGS01,LVS02},
to mention a few.
This may be explained by numerous applications of polypropylene 
in industry (ranged from oriented films for packaging to nonwoven fabrics
and reinforcing fibres).

Isotactic polypropylene is a semi-crystalline polymer containing
three different crystallographic forms \cite{IS00}:
monoclinic $\alpha$ crystallites,
hexagonal $\beta$ structures,
orthorhombic $\gamma$ polymorphs,
and ``smectic" mesophase (arrays of chains with a better order
in the longitudinal than in transverse chain direction).
At rapid cooling of the melt (which is typical of the injection-molding process),
$\alpha$ crystallites and smectic mesophase are mainly developed,
whereas $\beta$ and $\gamma$ polymorphs are observed as minority 
components \cite{KB97}.

The characteristic size of $\alpha$ spherulites in injection-molded 
specimens is estimated as 100 to 200 $\mu$m \cite{CCG98,KB97}.
These spherulites consist of crystalline lamellae with thickness 
of 10 to 20 nm \cite{CCG98,MHY00}.
A unique feature of $\alpha$ spherulites in iPP is the 
lamellar cross-hatching: 
development of transverse lamellae oriented in the direction 
perpendicular to the direction of radial lamellae in 
spherulites \cite{IS00,MHY00}.

The amorphous phase is located  (i) between spherulites,
(ii) in ``liquid pockets" \cite{VMH96} between lamellar stacks
inside spherulites, and (iii) between lamellae in lamellar stacks.
It consists of (i) mobile chains between spherulites, 
in liquid pockets and between radial lamellae inside lamellar stacks,
and (ii) chains with restricted mobility (the so-called 
``rigid amorphous fraction" \cite{VMH96})
in regions bounded by radial and tangential lamellae.

Stretching of iPP specimens induces inter-lamellar separation,
rotation and twist of lamellae,
fine and coarse slip of lamellar blocks and
their fragmentation \cite{AGU95,SSE99a},
chain slip through the crystals, 
sliding and separation of tie chains \cite{NT99,NT00},
and activation of the rigid amorphous fraction induced
by disintegration of transverse lamellae.
At large strains, these transformations result in cavitation, 
formation of fibrills and stress-induced 
crystallization \cite{ZBC99}.

Annealing and isothermal crystallization of iPP have attracted
an essential attention in the past five years,
see, e.g., \cite{IS00,MHY00,RQS98,YMT98,XSM98,ABM99,GHT02}.
Dramatic changes are observed in DSC (differential scanning
calorimetry) traces of isotactic polypropylene driven by
annealing at elevated temperatures (in the range from 110
to 170 $^{\circ}$C).
It is found that annealing in the low-temperature interval
(between 110 and 150 $^{\circ}$C) results in (i) a monotonical
increase in the melting peak \cite{IS00} and (ii) formation
of a broad low-temperature shoulder (second endotherm)
on a melting curve;
the intensity of this shoulder grows with annealing 
temperature \cite{ABM99}.
Annealing in the high-temperature interval (between 150
and 170 $^{\circ}$C) causes transformation of the second
endotherm into the main peak \cite{ABM99},
which may be attributed to a second-order phase
transition in the crystalline phase \cite{GHT02}.
The critical temperature corresponding to this transition
lies in the region between 157 \cite{MHY00} and 159 $^{\circ}$C
\cite{GHT02}.
The morphological analysis reveals that transformations of
the melting curves caused by annealing are accompanied by
changes in the crystalline structure (a pronounced reduction
in the level of cross-hatching with an increase in the
annealing temperature).

Although the effect of annealing on the micro-structure of crystallites
has been studied in detail, a little is known about the influence of
thermal treatment on the mechanical behavior of iPP.
Our studies on the viscoelastic response of isotactic polypropylene
annealed in the low-temperature interval (between 110 and 130 $^{\circ}$C)
revealed that the relaxation process is strongly affected
by annealing \cite{DC02a}.
The aim of the present work is to evaluate the effect of thermal 
treatment on the elastoplastic response of iPP.

An important shortcoming of conventional elastoplasticity theories
for solid polymers, see, e.g., \cite{BPA88,HB95,BK95,QPR97,SK98}, 
is that they are grounded on the concept of yield surface.
According to that approach, below the yield strain $\epsilon_{\rm y}$
(which is associated with the point of maximum on a stress--strain 
curve), a specimen demonstrates the nonlinear elastic 
(or viscoelastic) response.
This implies that the loading and unloading paths of stress--strain 
diagrams should coincide (provided that the strain 
rate is so large that the stress relaxation phenomenon
may be disregarded).
The experimental data presented in Section 2 show that this
assumption is inapplicable to isotactic polypropylene,
whose stress--strain curves for loading and unloading noticeably
differ from each other even when the maximal strain in a cyclic test
is far below the yield strain.

The objective of this paper is to develop constitutive equations
for the elastoplastic behavior of a semicrystalline polymer
that do not employ the concept of yield surface
and to apply these relations in fit observations in tensile cyclic 
tests with maximal strains in the sub-yield interval of deformations.

To make a constitutive model tractable from the mathematical standpoint,
we treat a semicrystalline polymer as an ensemble of meso-regions (MRs).
A meso-domain is thought of as an equivalent network of macromolecules 
bridged by junctions.
The mechanical behavior of a MR is associated with that of 
the amorphous phase, 
whereas the links between meso-domains that transmit the macro-strain
to individual MRs reflect the responses of crystallites.

A meso-region is treated as a linear elastic medium,
whereas deviations of the stress--strain curves in tensile tests 
from straight lines (that describe the response of a linear elastic solid)
are attributed to sliding of junctions between chains 
with respect to their reference positions in a stress-free material.
This is an essential simplification of the mechanism of micro-deformations,
which is tantamount to the assumption that only the amorphous phase 
is altered under active loading,
whereas the crystalline phase remains unchanged.
Elastic deformation of spherulites, inter-lamellar separation
and fine slip of lamellar blocks are not introduced into 
the model explicitly, but are taken into account implicitly in terms 
of ``average" parameters that characterize sliding of junctions in 
meso-domains.
This ``generalized" sliding process is described by a plastic strain,
$\epsilon_{\rm p1}$, whose rate of growth is assumed to be proportional 
to the rate of straining.

At unloading, junctions between chains in MRs move back 
to their initial positions with a decrease in the macro-strain $\epsilon$.
Because deformation of a semicrystalline polymer
results in micro-fracture of crystallites, the plastic strain $\epsilon_{\rm p1}$ 
(that decreases together with the macro-strain $\epsilon$)
is not sufficient to describe morphological transformations under
unloading.
To account for coarse slip and disintegration of lamellar blocks at unloading,
another plastic strain, $\epsilon_{\rm p2}$, is introduced,
which increases in time with a rate proportional to the rate of work
of external forces.

A similarity may be noted between splitting the plastic
strain, $\epsilon_{\rm p}$, into two components, 
$\epsilon_{\rm p1}$ and $\epsilon_{\rm p2}$, 
where the latter quantity is altered at unloading only, 
and a phenomenological model for the Mullins effect in particle-reinforced
elastomers \cite{OR99}.
According to the Ogden--Roxburgh concept, the difference between 
the stress--strain curves at active deformation and unloading 
is attributed to some damage parameter that changes along
the unloading path of a deformation history only.
An advantage of our model compared to the Ogden--Roxburgh 
approach is that the plastic strain, $\epsilon_{\rm p2}$, obeys a conventional
flow rule in elastoplasticity, whereas the damage parameter introduced 
in \cite{OR99} is governed by a kinetic equation whose physical meaning 
is not clear for semicrystalline polymers.

The exposition is organized as follows.
Observations in uniaxial loading--unloading tensile tests
are reported in Section 2.
Kinetic equations for the plastic strains, $\epsilon_{\rm p1}$ and
$\epsilon_{\rm p2}$, are introduced in Section 3.
Constitutive equations for a semicrystalline polymer at
isothermal uniaxial loading are derived in Section 4.
Adjustable parameters in the stress--strain relations are determined
in Section 5 by matching the experimental data.
A brief discussion of our findings is presented in Section 6.
Some concluding remarks are formulated in Section 7.

\section{Experimental procedure}

Isotactic polypropylene (Novolen 1100L) was supplied by BASF (Targor).
ASTM dumbbell specimens were injection molded 
with length 148 mm, width 10 mm and thickness 3.8 mm.
Uniaxial tensile tests were performed at room temperature 
on a testing machine Instron--5568 equipped with electro-mechanical 
sensors for the control of longitudinal strains 
in the active zone of samples.
The tensile force was measured by a standard load cell.
The longitudinal stress, $\sigma$, was determined
as the ratio of the axial force to the cross-sectional area
of stress-free specimens.

Four series of experiments were performed.
In the first series, the specimens were used as produced.
In the other series, the samples were annealed prior to testing
for 24 h at the temperatures 120, 140 and 160 $^{\circ}$C 
and slowly cooled by air.
To minimize the effect of physical aging, mechanical tests were 
carried out at least one day after thermal treatment.

In any test, a specimen was stretched with a cross-head speed 
of 10 mm/min (which corresponds to the Hencky strain rate 
$\dot{\epsilon}_{H}=2.09\cdot 10^{-4}$ s$^{-1}$)
up to the maximal strain, $\epsilon_{\max}$, and unloaded
with the same cross-head speed to the zero stress.
The chosen cross-head speed ensures nearly isothermal 
experimental conditions \cite{ITG02}, on the one hand, 
and it is sufficiently large to disregard the viscoelastic effects,
on the other (the maximal duration of a loading test does not
exceed 1 min).

Any series of experiments consisted of 6 tests with the
maximal strains $\epsilon_{\max}=0.02$, 0.04, 0.06, 0.08, 0.10 and 0.12.
Each test was performed on a new specimen.
The interval of strains under consideration includes the entire
sub-yield region (the yield strain for a non-annealed iPP 
$\epsilon_{\rm y}=0.13$, according to the supplier).
We confine ourselves to relatively small deformations,
because the purpose of the study is to approximate experimental data 
by using constitutive equations with small strains.

For the sake of brevity, we present only the stress--strain curves
for specimens annealed at 140 $^{\circ}$C.
The engineering stresses, $\sigma$, is plotted versus 
the engineering strain, $\epsilon$, in Figures 1 to 6.
The stress--strain diagrams for non-annealed specimens,
as well as for samples annealed at 120 and 160 $^{\circ}$C
have a similar shape.
The following features of the experimental curves 
are worth to be mentioned:
\begin{enumerate}
\item 
at all maximal strains, $\epsilon_{\max}$, the stress--strain curves
corresponding to the loading and unloading paths
substantially differ from each other,

\item
the unloading curves are strongly nonlinear,

\item
the residual strain (measured at the instant when the stress vanishes) 
noticeably increases with the maximal strain, $\epsilon_{\max}$.
\end{enumerate}

\section{A micro-mechanical model}

A semicrystalline polymer is treated as an equivalent network of chains.
The network is assumed to be strongly heterogeneous.
This heterogeneity is attributed to an inhomogeneity of interactions between 
chains in the amorphous phase and crystalline lamellae with various lengths 
and thicknesses.
The network is modelled as an ensemble of meso-regions (MRs)
with arbitrary shapes.
The characteristic length of a MR substantially exceeds the radius of
gyration for a macromolecule, and it is noticeably less than a size of
a sample.

Deformation of a specimen induces two processes in the network:
\begin{enumerate}
\item
Sliding of junctions (physical cross-links and entanglements that
bridge chains in the network) with respect to their positions in 
the stress-free state.

\item
Sliding of meso-domains in the ensemble with respect to each other.
\end{enumerate}
Sliding of junctions describes non-affine deformation of the network.
This process is determined by a plastic strain $\epsilon_{\rm p1}$.
Sliding of MRs with respect to each other is described
by a plastic strain $\epsilon_{\rm p2}$.
In a semicrystalline polymer,
the strain $\epsilon_{\rm p1}$ reflects sliding of junctions in
the amorphous phase, 
slippage of tie chains and fine slip of lamellar blocks.
The strain $\epsilon_{\rm p2}$ characterizes coarse slip
of lamellar blocks and their disintegration.

The total plastic strain, $\epsilon_{\rm p}$, equals the sum of 
the plastic strains driven by sliding of nodes and 
mutual displacements of meso-domains, 
\begin{equation}
\epsilon_{\rm p}=\epsilon_{\rm p1}+\epsilon_{\rm p2}.
\end{equation}
We suppose that meso-domains are connected by links
that transmit the macro-strain, $\epsilon$, to individual MRs.
This implies the conventional hypothesis that the macro-strain, $\epsilon$,
equals the sum of the elastic strain in meso-regions, $\epsilon_{\rm e}$, 
and the plastic strain, $\epsilon_{\rm p}$,
\begin{equation}
\epsilon=\epsilon_{\rm e}+\epsilon_{\rm p}.
\end{equation}
To simplify the analysis, we assume that the elastic strains in various MRs
coincide.

Deformation of a specimen results in evolution of the plastic strain, 
$\epsilon_{\rm p1}$, both at the stages of active loading and unloading.
We suppose that the rate of changes in the plastic strain, $\epsilon_{\rm p1}$, 
is proportional to the rate of changes in the macro-strain, $\epsilon$,
\begin{equation}
\frac{d\epsilon_{\rm p1}}{dt} (t)=\varphi (\epsilon_{\rm e}(t)) \frac{d\epsilon}{dt}(t),
\end{equation}
where the coefficient of proportionality, $\varphi$, depends on
the elastic strain $\epsilon_{\rm e}$.
It is assumed that the function $\varphi(\epsilon_{\rm e})$
vanishes at the zero elastic strain, $\varphi(0) =0$,
monotonically increases with $\epsilon_{\rm e}$,
and tends to some constant $a\in [0,1]$ for relatively large elastic strains,
\[
\lim_{\epsilon_{\rm e}\to \infty} \varphi(\epsilon_{\rm e}) =a,
\]
where $a$ is the rate of sliding of junctions in MRs corresponding 
to a developed plastic flow.
The inequality $a\geq 0$ means that junctions slide in the direction
that is determined by the macro-strain $\epsilon$.
The condition $a\leq 1$ ensures that the rate of sliding does not exceed
the rate of straining.

To approximate experimental data, we employ the function
\begin{equation}
\varphi(\epsilon_{\rm e})=a \Bigl [ 1-\exp\Bigl (-\frac{\epsilon_{\rm e}}{\varepsilon}\Bigr )
\Bigr ].
\end{equation}
This function is determined by two adjustable parameters, 
$a$ and $\varepsilon$,
where the quantity $\varepsilon$ characterizes how ``large" 
an elastic strain, $\epsilon_{\rm e}$, is.

With reference to \cite{AGU95,SSE99a}, we suppose that under
active loading in the sub-yield region, lamellar fragmentation
does not occur and the plastic strain $\epsilon_{\rm p2}$ vanishes,
\begin{equation}
\frac{d\epsilon_{\rm p2}}{dt}(t)=0
\qquad 
\Bigl ( \frac{d\epsilon}{dt}(t)\geq 0,
\quad \sigma(t)\geq 0 \Bigr ).
\end{equation}
According to Eq. (5), the entire dissipation of energy
at active deformation is attributed to sliding of junctions 
with respect to their reference positions,
whereas displacements of MRs with respect to each other 
are disregarded.

It is assumed that under unloading,
meso-domains slide with respect to each other 
as they are driven by a positive macro-stress $\sigma$.
These mutual displacements of MRs are characterized 
by a plastic strain, $\epsilon_{\rm p2}$, that grows with time $t$.
An increase in $\epsilon_{\rm p2}$ during unloading reflects
coarse slip and fragmentation of deformed lamellae.
The evolution of the plastic strain, $\epsilon_{\rm p2}$, 
is described by the flow rule
\begin{equation}
\frac{d\epsilon_{\rm p2}}{dt}(t)=-K \sigma(t) \frac{d\epsilon}{dt}(t)
\qquad 
\Bigl ( \frac{d\epsilon}{dt}(t) <0,
\quad \sigma(t)\geq 0 \Bigr ).
\end{equation}
Equation (6) means that the rate of changes in $\epsilon_{\rm p2}$ is
proportional to the rate of work of external loads
\[ 
J(t)=-\sigma(t) \frac{d\epsilon}{dt}(t),
\]
where the sign ``$-$" accounts for the opposite directions of
the longitudinal stress and the strain increment.
With reference to \cite{Mie95}, we suppose that
the coefficient of proportionality, $K\geq 0$, is a function of
the maximal plastic strain, $\epsilon_{\rm p1}$, reached:
$K=K(\epsilon_{\rm p1}^{\circ})$, where
\[
\epsilon_{\rm p1}^{\circ}(t)=\max_{0\leq \tau\leq t} \epsilon_{\rm p1}(\tau).
\]
To match observations, we use the linear function
\begin{equation}
K=K_{0}+K_{1}\epsilon_{\rm p1}^{\circ},
\end{equation}
where $K_{m}$ ($m=0,1$) are adjustable parameters.
Equation (7) means that the larger is the plastic strain, $\epsilon_{\rm p1}$, 
driven by fine slip of lamellar blocks under active loading, 
the higher is the rate of the plastic strain, $\epsilon_{\rm p2}$,
that reflects coarse slip and fragmentation of lamellae at unloading.

Separation of tie chains from lamellae
and disintegration of lamellar blocks result in a decrease
in the number of MRs to which the macro-strain, $\epsilon$,
is transmitted by surrounding meso-domains.
This decrease is attributed to two different processes:
(i) mechanically-induced separation of individual MRs from the ensemble,
and (ii) screening of meso-domains by stacks of disintegrated lamellae.

To describe evolution of an ensemble of meso-domains, 
we introduce the average number of MRs, $N_{0}$, 
per unit mass of a virgin specimen (where all MRs are 
connected with one another) and the average number of MRs,
$N(t)$, in the deformed specimen at time $t\geq 0$
(where some meso-regions are separated from the ensemble).

At the stage of active loading in the sub-yield
region, separation of MRs from the ensemble does not take place,
which implies that $N(t)$ remains constant,
\begin{equation}
\frac{dN}{dt}(t)=0
\qquad  
\Bigl ( \frac{d\epsilon}{dt}(t)\geq 0,
\quad \sigma(t)\geq 0 \Bigr ),
\qquad
N(0)=N_{0}.
\end{equation}
Changes in the function $N(t)$ at unloading are governed 
by a first order kinetic equation, according to which
the relative number of MRs separated from the ensemble
per unit time is proportional to the increment of the plastic strain, 
$\epsilon_{\rm p2}$, 
\begin{equation}
-\frac{1}{N(t)} \frac{dN}{dt}(t)=\kappa \frac{d\epsilon_{\rm p2}}{dt}(t) 
\qquad  
\Bigl ( \frac{d\epsilon}{dt}(t) <0,
\quad \sigma(t)\geq 0 \Bigr ).
\end{equation}
By analogy with Eq. (6), the coefficient $\kappa >0$ is treated
as a function of the maximal plastic strain $\epsilon_{\rm p1}^{\circ}$:
$\kappa=\kappa(\epsilon_{\rm p1}^{\circ})$.
The pre-factor $\kappa$ in Eq. (9) is assumed to be rather large
for small plastic strains, $\epsilon_{\rm p1}$, i.e., when straining
of a specimen produces no fine slip of lamellar blocks,
and it monotonically decreases with $\epsilon_{\rm p1}$.
This decrease is attributed to alignment of lamellar blocks
driven by their fine slip, which, in turn, diminishes the
probability of formation of disordered lamellar stacks 
at unloading.
These stacks do not transmit the macro-strain, $\epsilon$, 
to the amorphous domains surrounded by them, which results
in isolation of these regions from the ensemble (in a way
silimar to the formation of regions of occluded rubber in
particle-reinforced elastomers \cite{WRC93}).

To fit experimental data, we adopt the function
\begin{equation}
\kappa=\kappa_{0}+\kappa_{1}\exp 
\Bigl (-\frac{\epsilon_{\rm p1}^{\circ}}{e_{\ast}}\Bigr ),
\end{equation}
where $\kappa_{m}$ ($m=0,1$) and $e_{\ast}$ are adjustable
parameters. 
Equation (10) implies that $\kappa$ monotonically decreases
from the initial value $\kappa_{0}+\kappa_{1}$ and tends
to the limiting value $\kappa_{0}$ at relatively large plastic
strains, $\epsilon_{\rm p1}$.
The quantity $e_{\ast}$ indicates how ``large" the maximal
plastic strain, $\epsilon_{\rm p1}^{\circ}$, is.

\section{Constitutive equations}

Under isothermal uniaxial deformation,
a meso-domain is treated as a linear elastic solid with the mechanical energy
\[
w=\frac{1}{2} \mu \epsilon_{\rm e}^{2},
\]
where the constant $\mu>0$ is the average rigidity of a MR.
Neglecting the energy of interaction between meso-regions, 
we calculate the strain energy density per unit mass of 
a polymer as the sum of the mechanical energies of MRs,
\begin{equation}
W(t)=\frac{\mu}{2} N(t)\epsilon_{\rm e}^{2}.
\end{equation}
It follows from Eqs. (1) to (3) and (11) that the derivative of the 
function $W$ with respect to time is given by
\begin{equation}
\frac{dW}{dt}(t) = \frac{\mu}{2} \frac{dN}{dt}(t)  \epsilon_{\rm e}^{2}(t)
+\mu N(t) \epsilon_{\rm e}(t) 
\Bigl [ \Bigl (1-\varphi(\epsilon_{\rm e}(t))\Bigr ) \frac{d\epsilon}{dt}(t)
-\frac{d\epsilon_{\rm p2}}{dt}(t) \Bigr ].
\end{equation}
The Clausius-Duhem inequality reads 
\[
Q(t)=-\frac{dW}{dt}(t)+\frac{1}{\rho}\sigma(t)\frac{d\epsilon}{dt}(t) \geq 0,
\]
where $\rho$ is mass density,
and $Q$ is internal dissipation per unit mass.
Substition of Eq. (12) into this formula results in
\begin{equation}
Q(t)= \frac{1}{\rho} \Bigl [ \sigma(t)-\rho \mu N(t)\epsilon_{\rm e}(t)
\Bigl (1-\varphi(\epsilon_{\rm e}(t))\Bigr ) \Bigr ] \frac{d\epsilon}{dt}(t) +Y(t)\geq 0,
\end{equation}
where
\begin{equation}
Y(t) = \mu N(t)\epsilon_{\rm e}(t) \Bigl [ \frac{d\epsilon_{\rm p2}}{dt}(t)
-\frac{1}{2N(t)} \frac{dN}{dt}(t)\epsilon_{\rm e}(t) \Bigr ].
\end{equation}
Equating the expression in square brackets in Eq. (13) to zero,
we find that
\begin{equation}
\sigma(t) =E(t) \epsilon_{\rm e}(t) \Bigl [1-\varphi(\epsilon_{\rm e}(t))\Bigr ]
\end{equation}
with
\begin{equation}
E(t)= \rho \mu N(t).
\end{equation}
Equations (5), (8) and (14) imply that the function $Y$ 
vanishes at active loading, 
\begin{equation}
Y(t)=0
\qquad  
\Bigl ( \frac{d\epsilon}{dt}(t)\geq 0,
\quad \sigma(t)\geq 0 \Bigr ).
\end{equation}
It follows from Eqs. (6), (9), and (14) to (16) that at unloading,
\begin{equation}
Y(t)=-\frac{K}{\rho} \Bigl ( 1+\frac{\kappa}{2}\Bigr )
\frac{\sigma^{2}(t)}{1-\varphi(\epsilon_{\rm e}(t))}\frac{d\epsilon}{dt}(t)
\qquad  
\Bigl ( \frac{d\epsilon}{dt}(t) < 0,
\quad \sigma(t)\geq 0 \Bigr ).
\end{equation}
Combining Eqs. (13), (15) and (16), we find that the internal dissipation 
per unit mass, $Q(t)$, coincides with $Y(t)$.
According to Eqs. (17) and (18), the function $Y(t)$ is non-negative.
This means that the Clausius--Duhem inequality is satisfied 
for an arbitrary deformation program, provided that the stress, 
$\sigma$, is given by Eq. (15).

It follows from Eqs. (8), (9) and (16) that the elastic modulus, $E$,
obeys the differential equations
\begin{eqnarray}
&& \frac{dE}{dt}(t) = 0,
\quad
E(0)=E_{0}
\qquad
\Bigl ( \frac{d\epsilon}{dt}(t)\geq 0,
\quad \sigma(t)\geq 0 \Bigr ),
\nonumber\\
&& \frac{1}{E(t)} \frac{dE}{dt}(t) =
-\kappa \frac{d\epsilon_{\rm p2}}{dt}(t) 
\qquad  
\Bigl ( \frac{d\epsilon}{dt}(t) <0,
\quad \sigma(t)\geq 0 \Bigr ),
\end{eqnarray}
where $E_{0}=\mu\rho N_{0}$ is the elastic modulus for 
a virgin specimen.

The assumption that elastic moduli are strongly affected 
by mechanical factors is widely used in elastoplasticity 
theories for geomaterials \cite{Hou85,RMW00}.
For granular media,
this effect is attributed to the growth of voids between particles 
under active loading and to the contractive response at unloading
(which means that elastic moduli become functions of
the volumetric strain \cite{RMW00}).
In the present model, the plastic strain, $\epsilon_{\rm p2}$, 
plays a role similar to that the mean elastic strain plays 
in elastoplasticity theories for granular materials: 
it characterizes the level of disorder produced 
by mechanically-induced lamellar fragmentation.
The difference between our approach and previous studies is
that for an arbitrary time-dependent loading program,
the elastic modulus, $E$, depends not on the current strain, 
but on the entire history of deformation [because Eqs. (19)
cannot be integrated explicitly when $\kappa$ is a function 
of the maximal plastic strain, $\epsilon_{\rm p1}^{\circ}$, 
reached under active loading].

Uniaxial deformation of a semi-crystalline polymer
is described by Eqs. (1), (2), (4) to (6),  (15) and (19).
Any stress--strain curve for cyclic loading
is determined by 5 adjustable parameters:
\begin{enumerate}
\item
the initial elastic modulus $E_{0}$,

\item
the rate of a developed plastic flow $a$,

\item
the strain $\varepsilon$ that characterizes transition to a steady
plastic flow,

\item
the rate of sliding of MRs with respect to each other
$K=K(\epsilon_{\rm p1}^{\circ})$, 

\item
the rate of separation of meso-domains from an ensemble 
$\kappa=\kappa(\epsilon_{\rm p1}^{\circ})$.
\end{enumerate}
This number is quite comparable with the number of material
constants in other constitutive models in elastoplasticity,
see, e.g., \cite{HB95,BK95,SK98}.
It should be noted, however, that most of these concepts 
fail to adequately describe the mechanical behavior of
polymers in cyclic tests \cite{BKR02}.

An important advantage of the stress--strain relations
(1), (2), (4) to (6),  (15) and (19) is that 3 constants, $E_{0}$, $a$
and $\varepsilon$, are found by fitting experimental data for
the loading path of a stress--strain curve,
whereas the other two parameters, $K$ and $\kappa$, are
determined by matching observations for the unloading path.

The dependences of $K$ and $\kappa$ 
on the maximal plastic strain, $\epsilon_{\rm p1}^{\circ}$,
are described by Eqs. (7) and (10).
To find the coefficients $K_{m}$, $\kappa_{m}$ and $e_{\ast}$ is these
equations, the quantities $K$ and $\kappa$ (determined by
matching observations in a series of cyclic tests with various
maximal strains $\epsilon_{\max}$) are approximated by
functions (7) and (10).

\section{Fitting of observations}

We begin with the approximation of the stress--strain diagrams 
for active loading.
It follows from Eqs. (1), (2), (4), (5), (15) and (19) that the 
stress, $\sigma$, is given by
\begin{equation}
\sigma(\epsilon) = E_{0} (\epsilon-\epsilon_{\rm p1})\Bigl \{ 1- a 
\Bigl [1-\exp \Bigl (-\frac{\epsilon-\epsilon_{\rm p1}}{\varepsilon}\Bigr )
\Bigr ]\Bigr \},
\end{equation}
where the plastic strain, $\epsilon_{\rm p1}$, satisfies the nonlinear 
differential equation
\begin{equation}
\frac{d\epsilon_{\rm p1}}{d\epsilon}(\epsilon)
=a \Bigl [1-\exp \Bigl (-\frac{\epsilon-\epsilon_{\rm p1}}{\varepsilon}\Bigr )
\Bigr ],
\qquad
\epsilon_{\rm p1}(0)=0.
\end{equation}
The loading path of any stress--strain curve is determined 
by 3 material constants: $E_{0}$, $a$ and $\varepsilon$.
To find these quantities, we fix some intervals $[0,a_{\max}]$ 
and $[0,\varepsilon_{\max}]$, where the ``best-fit" parameters 
$a$ and $\varepsilon$ are assumed to be located,
and divide these intervals into $J$ subintervals by
the points $a_{i}=i\Delta a$ and $\varepsilon_{j}=j\Delta \varepsilon$  
($i,j=1,\ldots,J$) with $\Delta a=a_{\max}/J$ and 
$\Delta \varepsilon=\varepsilon_{\max}/J$.
For any pair, $\{ a_{i}, \varepsilon_{j} \}$, 
Eq. (21) is integrated numerically by the Runge--Kutta method
with the step $\Delta \epsilon=1.0\cdot 10^{-5}$.
Given a pair, $\{ a_{i}, \epsilon_{j} \}$, the elastic modulus, 
$E_{0}=E_{0}(i,j)$, is found by the least-squares method
from the condition of minimum of the function
\[
F(i,j)=\sum_{\epsilon_{m}} \Bigl [ \sigma_{\rm exp}(\epsilon_{m})
-\sigma_{\rm num}(\epsilon_{m}) \Bigr ]^{2},
\]
where the sum is calculated over all experimental points,
$\epsilon_{m}$, on a loading path, 
$\sigma_{\rm exp}$ is the stress measured in a tensile test, 
and $\sigma_{\rm num}$ is given by Eq. (20).
The ``best-fit" parameters $a$ and $\varepsilon$ are determined
from the condition of minimum of the function $F$ 
on the set $ \{ a_{i}, \varepsilon_{j} \quad (i,j=1,\ldots, J)  \}$.

The material constants $E_{0}$, $a$ and $\varepsilon$ that
minimize the discrepancies between the experimental data
and the results of numerical analysis are found for any
stress--strain curve independently.
Afterwards, we calculate the average values of these quantities 
and their standard deviations for 4 series of tests.
These values are presented in Table 1 for non-annealed specimens
and for samples annealed at 120, 140 and 160 $^{\circ}$C.
The table shows that the elastic modulus, $E_{0}$, 
and the rate of developed plastic flow, $a$, are determined
with a high level of accuracy (the relative deviations are less 
than 8 and 13 \%, respectively), whereas the accuracy in
determining the strain, $\varepsilon$, that characterizes transition
to a steady-state plastic flow, is rather low (about 29 \%).

To find the quantities $K$ and $\kappa$,  we approximate 
the unloading paths of the stress--strain curves.
It follows from Eqs. (1), (2), (4), (6), (15) and (18) that the stress, $\sigma$, 
is given by
\begin{equation}
\sigma(\epsilon) = E(\epsilon_{\rm p2})(\epsilon-\epsilon_{\rm p1}
-\epsilon_{\rm p2})\Bigl \{ 1-a  
\Bigl [ 1 -\exp\Bigl (-\frac{\epsilon-\epsilon_{\rm p1}-\epsilon_{\rm p2}}{\varepsilon}
\Bigr )\Bigr ]\Bigr \},
\end{equation}
where the elastic modulus, $E$, reads
\begin{equation}
E(\epsilon_{\rm p2})=E_{0}\exp (-\kappa \epsilon_{\rm p2}).
\end{equation}
The plastic strains, $\epsilon_{\rm p1}$ and $\epsilon_{\rm p2}$,
obey the differential equations
\begin{equation}
\frac{d\epsilon_{\rm p1}}{d\epsilon}(\epsilon) = 
a \Bigl [ 1 -\exp\Bigl (-\frac{\epsilon-\epsilon_{\rm p1}
-\epsilon_{\rm p2}}{\varepsilon}\Bigr )\Bigr ],
\qquad
\frac{d\epsilon_{\rm p2}}{d\epsilon}(\epsilon) =
-K \sigma(\epsilon)
\end{equation}
with the ``initial" conditions 
\[
\epsilon_{\rm p1}(\epsilon_{\max})=\epsilon_{\rm p1}^{\circ},
\qquad
\epsilon_{\rm p2}(\epsilon_{\max})=0.
\]
To approximate experimental data in a cyclic test with 
a maximal tensile strain $\epsilon_{\max}$, 
we apply an algorithm similar to that used to match 
the stress--strain curves at active loading.
We fix some intervals $[0,K_{\max}]$ and $[0,\kappa_{\max}]$, 
where the ``best-fit" parameters $K$ and $\kappa$ are assumed to be located,
and divide these intervals into $J$ subintervals by
the points $K_{i}=i\Delta K$ and $\kappa_{j}=j\Delta \kappa$  ($i,j=1,\ldots,J$)
with $\Delta K=K_{\max}/J$ and $\Delta \kappa=\kappa_{\max}/J$.
For any pair, $\{ K_{i}, \kappa_{j} \}$, 
Eqs. (20) to (22) are integrated numerically by the Runge--Kutta method
with the step $\Delta \epsilon=1.0\cdot 10^{-5}$
and with the value of $E_{0}$ found by fitting the loading paths
of the stress--strain curves.
The ``best-fit" parameters $K$ and $\kappa$ are determined from the
condition of minimum of the function $F$ 
on the set $ \{ K_{i}, \kappa_{j} \quad (i,j=1,\ldots, J)  \}$.

The parameter $K$ is plotted versus $\epsilon_{\rm p1}^{\circ}$ in 
Figure 7.
The maximal plastic strain, $\epsilon_{\rm p1}^{\circ}$, 
is determined by numerical integration of Eq. (21) 
from $\epsilon=0$ to $\epsilon=\epsilon_{\max}$.
The experimental data are fitted by Eq. (7),
where the coefficients $K_{m}$ ($m=0,1$) are found by
the least-squares technique.

The quantity $\kappa$ is depicted as a function of
$\epsilon_{\rm p1}^{\circ}$ in Figure 8 together with 
its approximation by Eq. (10).
The strain $e_{\ast}$ is determined
by the steepest-descent method, whereas the coefficients
$\kappa_{m}$ ($m=0,1$) are calculated by the least-squares
algorithm.
Unlike Figure 7, where the only curve provides an acceptable
approximation of all experimental data,
adjustable parameters in Eq. (10) are found independently
for specimens not-subjected to thermal treatment
and annealed at temperatures in the low-temperature 
region, on the one hand, and for samples annealed at 160 $^{\circ}$C,
on the other.

\section{Discussion}

Figure 1 to 6 demonstrate fair agreement between the observations
in loading--unloading tests with various maximal strains, $\epsilon_{\max}$,
and the results of numerical simulation.
The same quality of fitting has been reached also for
the stress--strain curves for specimens not subjected
to thermal treatment and those annealed at 120 and 140 $^{\circ}$C.
These figures confirm that the model correctly describes
the mechanical response of semicrystalline polymers in cyclic tests.

Table 1 shows that annealing of iPP in the low-temperature region
does not affect the elastic modulus, $E_{0}$, and the rate of developed
plastic flow, $a$.
Annealing at 160 $^{\circ}$C causes a noticeable
increase (from 20 to 30 \%) in the elastic modulus, $E_{0}$,
and a growth of the rate of steady plastic flow, $a$, that reaches its
ultimate value.
Although the increase in $a$ is not rather high (about 6 to 14 \%), 
the fact that the rate
of developed plastic flow reaches unity means that at relatively
large elongations, the elastic strain, $\epsilon_{\rm e}$, does
not grow,  
and an increase in the macro-strain, $\epsilon$, 
is totally compensated by the same increase in the plastic strain, 
$\epsilon_{\rm p1}$.
On the contrary, the rate of developed plastic flow in specimens 
not subjected to thermal treatment, as well as in those
annealed in the low-temperature region is noticeably less than unity,
which implies a strong increase in the elastic strain under
stretching.
This difference becomes important for the analysis of stress--strain curves 
at large elongations (appropriate data are not presented):
necking of specimens not subjected to annealing occurs
at the Hencky strains about 0.2, whereas necking of specimens 
annealed in the high-temperature region is not observed at
the Hencky strains up to 0.6 .

The growth of the elastic modulus, $E_{0}$, of specimens annealed at 160 $^{\circ}$C
may be attributed to an increase in perfectness of crystalline
lamellae (associated with transition from a statistically disordered $\alpha_{1}$ phase 
into an ordered $\alpha_{2}$ phase characterized by regularity
in the up and down positions of methyl groups along the chains \cite{GHT02}).
An increase in the rate of developed plastic flow, $a$,
may be ascribed to a decrease in the level of cross-hatching in $\alpha$
spherulites with the growth of annealing temperature \cite{YMT98} 
and the total disappearance of transverse lamellae above 
the critical temperature $T_{\rm c}=159$ $^{\circ}$C \cite{GHT02}.

Table 1 shows that the parameter $\varepsilon$ is independent of the
annealing temperature (within the range of experimental uncertainities).
This conclusion appears to be quite natural, because $\varepsilon$ 
characterizes sliding of junctions between chains in MRs.
This quantity describes the response of the amorphous
phase that is not affected by thermal treatment.
The value of $\varepsilon$ (in the range between 3 and 4 \%) found by
fitting the stress--strain curves appears to be rather close to 
a critical strain, $\epsilon_{\rm c}$, determined by matching relaxation 
curves \cite{DC02b} as the strain at which the rate of relaxation 
becomes independent of mechanical factors.

Figure 7 reveals that the parameter $K$ is not affected by 
annealing temperature.
At first glance, this result seems rather surprising, because $K$
is responsible for mutual displacements of MRs
driven by coarse slip and fragmentation of lamellar blocks.
Two explanations may be provided for our findings.
The first is based on the Nitta--Takayanagi \cite{NT99}
and Meyer--Pruitt \cite{MP01} concepts that attribute
plastic deformation of semicrystalline polymers
to slippage of tie chains along lamellae and pulling out of 
chains from disintegrated lamellar blocks (both processes
are associated with transformations in the amorphous phase
whose state is not influenced by thermal treatment).
The other explanation is based on the assumption that
an increase in perfectness of crystallites at annealing (which makes
lamellae stronger) is compensated by disappearance of
tangential lamellae (which results in weakening of spherulites).
Because these morphological transformations in iPP
lead to opposite changes in the mechanical response, their
combined effect may be negligible.

According to Figure 8, the parameter $\kappa$ is not affected
by annealing in the low-temperature region, but is strongly
influenced by high-temperature annealing.
The limiting value of this quantity, $\kappa_{0}$, that corresponds
to a developed plastic flow at unloading, increases approximately
by twice after annealing at 160 $^{\circ}$C.
Because $\kappa$ characterizes changes in the elastic modulus
driven by lamellar fragmentation, see Eq. (19), we associate
this growth with weakening of spherulites induced by disappearance
of cross-hatching.

The strain, $e_{\ast}$, that determines transition to a steady
plastic flow at unloading (for all specimens, but those annealed
at 160 $^{\circ}$C) is very close to the strain, $\varepsilon$,
that characterizes transition to a developed plastic flow at
active loading.
This implies an opportunity to reduce the number of adjustable
parameters in the constitutive equations by replacing $e_{\ast}$
by $\varepsilon$.
However, an additional analysis of the physical basis for 
this simplification is required.

\section{Concluding remarks}

Four series of tensile loading--unloading tests have been performed
on injection-molded isotactic polypropylene at room temperature.
In the first series, specimens were used as produced,
whereas in the other series, samples were annealed for 24 h
at 120, 140 and 160 $^{\circ}$C prior to testing.

Constitutive equations have been derived for the elastoplastic response 
of semicrystalline polymers.
A polymer is treated as an equivalent network of chains bridged
by junctions.
The network is assumed to be strongly heterogeneous, 
and it is modelled as an ensemble of meso-regions linked 
with each other.
Under active loading in the sub-yield region of deformations, 
junctions between chains in MRs slide with respect to their 
reference positions (which describes sliding of tie chains
and fine slip of lamellar blocks).
At unloading, this non-affine deformation of the network
is accompanied by displacements of meso-domains with respect to
each other (which describes coarse slip and fragmentation 
of lamellar blocks).
Destructure of lamellae results in separation of some MRs 
from the ensemble (driven by breakage
of links between isolated meso-domains and the network
and screening of macro-strain in these meso-domains by
surrounding stacks of disintegrated lamellae.

Kinetic equations are proposed for the rates of two plastic strains 
that describe these processes and for the rate of
separation of MRs from an ensemble.
Stress--strain relations for isothermal uniaxial deformation 
are developed by using the laws of thermodynamics.
The constitutive equations are determined by 5 adjustable 
parameters that are found by fitting the experimental data.
Fair agreement is demonstrated between the experimental 
stress--strain curves and the results of numerical simulation.

The following conclusions are drawn:
\begin{enumerate}
\item
Annealing in the low-temperature region does not affect
the material constants that reflect the elastoplastic response
of iPP.

\item
Annealing in the high-temperature region results in an increase
in the elastic modulus $E_{0}$ (which is attributed to the growth of
the perfectness of crystals induced by $\alpha_{1}\to\alpha_{2}$
transition).

\item
Annealing in the high-temperature region causes the growth of
the rate of developed plastic flow $a$ (that reaches its ultimate
value) and an increase in the rate of separation of meso-regions 
from an ensemble $\kappa$ (these changes are associated 
with disappearance of transversal lamellae).

\item
The rate of plastic strain, $K$, linearly grows with
the maximal plastic strain, $\epsilon_{\rm p1}^{\circ}$,
which means that sliding of junctions in MRs under active loading
activates coarse slip and fragmentation of lamellae at
unloading.
The activation process appears to be independent of 
the perfectness of crystallites.

\item
The strains $\varepsilon$ and $e_{\ast}$ that characterize transitions
to steady plastic flows at active loading and unloading, respectively,
are rather close to each other and are weakly affected by thermal
treatment.
This result confirms the hypothesis \cite{NT99,MP01} that 
elastoplastic deformation of a semicrystalline polymer
in a sub-yield region is mainly associated with transformations 
in the amorphous phase.
\end{enumerate}

\newpage

\section*{List of figures}
\parindent 0 mm

{\bf Figure 1:}
The stress $\sigma$ MPa versus strain $\epsilon$ 
in a tensile loading--unloading test with the maximum strain
$\epsilon_{\max}=0.02$.
Circles: experimental data on a specimen annealed
at $T=140$ $^{\circ}$C.
Solid line: results of numerical simulation
\vspace*{2 mm}

{\bf Figure 2:}
The stress $\sigma$ MPa versus strain $\epsilon$ 
in a tensile loading--unloading test with the maximum strain
$\epsilon_{\max}=0.04$.
Circles: experimental data on a specimen annealed
at $T=140$ $^{\circ}$C.
Solid line: results of numerical simulation
\vspace*{2 mm}

{\bf Figure 3:}
The stress $\sigma$ MPa versus strain $\epsilon$ 
in a tensile loading--unloading test with the maximum strain
$\epsilon_{\max}=0.06$.
Circles: experimental data on a specimen annealed
at $T=140$ $^{\circ}$C.
Solid line: results of numerical simulation
\vspace*{2 mm}

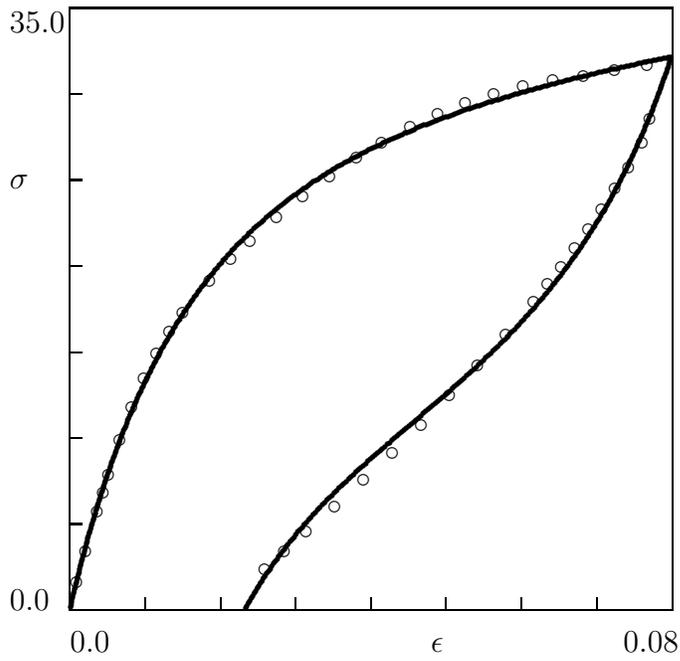
{\bf Figure 4:}
The stress $\sigma$ MPa versus strain $\epsilon$ 
in a tensile loading--unloading test with the maximum strain
$\epsilon_{\max}=0.08$.
Circles: experimental data on a specimen annealed
at $T=140$ $^{\circ}$C.
Solid line: results of numerical simulation
\vspace*{2 mm}

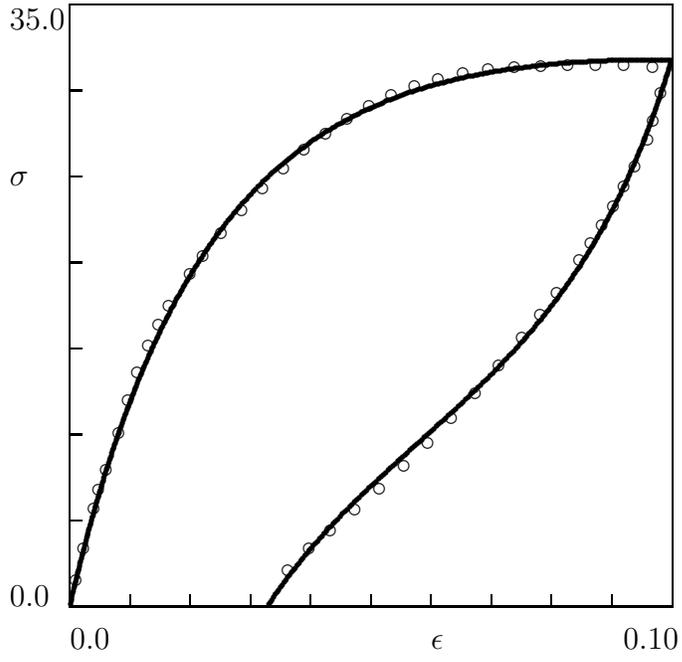
{\bf Figure 5:}
The stress $\sigma$ MPa versus strain $\epsilon$ 
in a tensile loading--unloading test with the maximum strain
$\epsilon_{\max}=0.10$.
Circles: experimental data on a specimen annealed
at $T=140$ $^{\circ}$C.
Solid line: results of numerical simulation
\vspace*{2 mm}

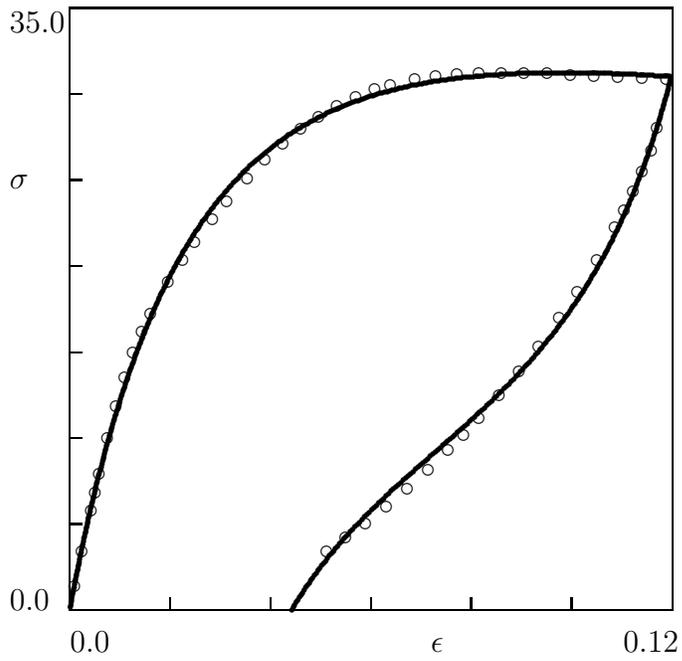
{\bf Figure 6:}
The stress $\sigma$ MPa versus strain $\epsilon$ 
in a tensile loading--unloading test with the maximum strain
$\epsilon_{\max}=0.12$.
Circles: experimental data on a specimen annealed
at $T=140$ $^{\circ}$C.
Solid line: results of numerical simulation
\vspace*{2 mm}

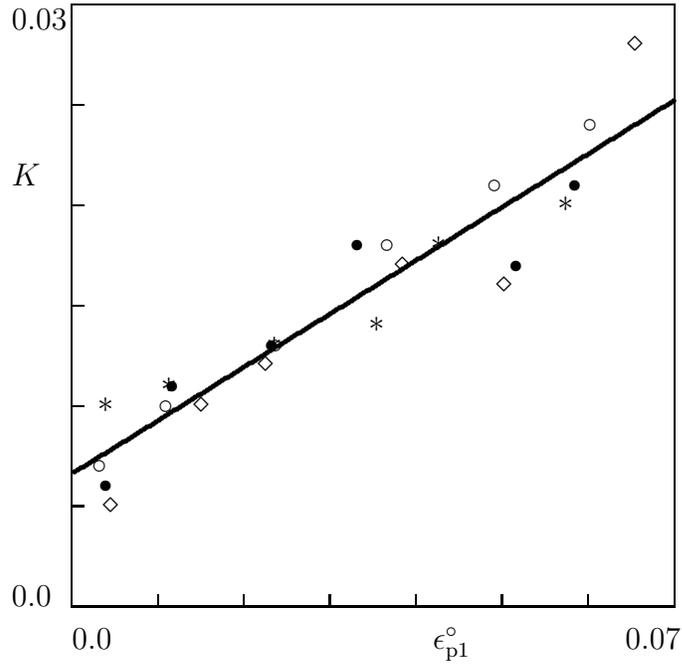
{\bf Figure 7:}
The parameter $K$ MPa$^{-1}$ versus the maximal
plastic strain $\epsilon_{\rm p1}^{\circ}$.
Symbols: treatment of observations.
Unfilled circles: non-annealed specimens;
filled circles: specimens annealed at $T=120$ $^{\circ}$C;
asterisks: specimens annealed at $T=140$ $^{\circ}$C;
diamonds: specimens annealed at $T=160$ $^{\circ}$C.
Solid line: approximation of the experimental data
by Eq. (7) with $K_{0}=0.0066$ and $K_{1}=0.2659$
\vspace*{2 mm}

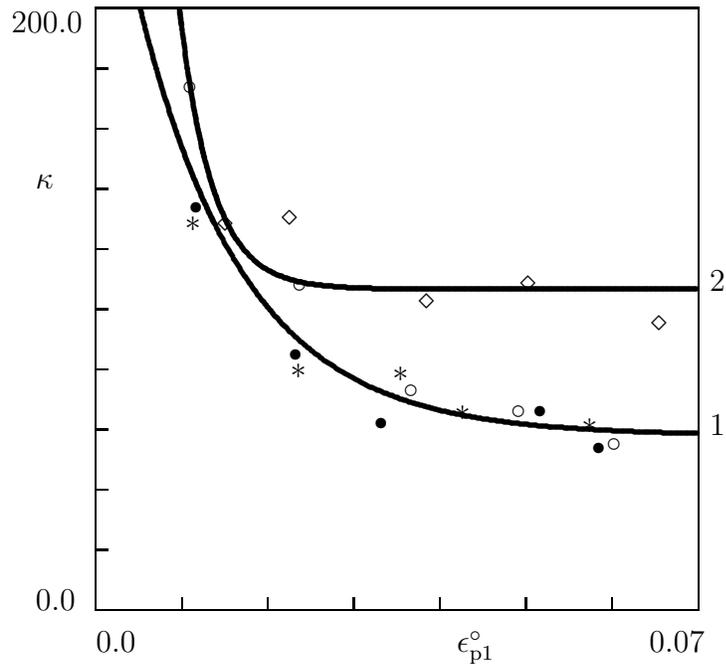
{\bf Figure 8:}
The dimensionless parameter $\kappa$ versus the maximal
plastic strain $\epsilon_{\rm p1}^{\circ}$.
Symbols: treatment of observations.
Unfilled circles: non-annealed specimens;
filled circles: specimens annealed at $T=120$ $^{\circ}$C;
asterisks: specimens annealed at $T=140$ $^{\circ}$C;
diamonds: specimens annealed at $T=160$ $^{\circ}$C.
Solid lines: approximation of the experimental data
by Eq. (10).
Curve 1: $\kappa_{0}=57.96$, $\kappa_{1}=215.83$, 
$e_{\ast}=1.23\cdot 10^{-2}$;
curve 2: $\kappa_{0}=106.68$, $\kappa_{1}=1184.71$, 
$e_{\ast}=3.83\cdot 10^{-3}$
\newpage

\begin{table}[tbh]
\begin{center}
\caption{Adjustable parameters $E_{0}$ GPa, $a$ and $\varepsilon$ 
for specimens annealed at various temperatures $T$ $^{\circ}$C
(the values in parentheses indicate standard deviations)}
\vspace*{6 mm}


\end{center}
\vspace*{5 mm}

\caption{}
\end{figure}
%\end{document}

%\documentclass[12pt]{article}
%\pagestyle{empty}
%\begin{document}
%\setlength{\unitlength}{1.0 mm}
\begin{figure}[tbh]
\begin{center}
%
\end{center}
\vspace*{5 mm}

\caption{}
\end{figure}
%\end{document}

%\documentclass[12pt]{article}
%\pagestyle{empty}
%\begin{document}
%\setlength{\unitlength}{1.0 mm}
\begin{figure}[tbh]
\begin{center}
%
\end{center}
\vspace*{5 mm}

\caption{}
\end{figure}
%\end{document}

%\documentclass[12pt]{article}
%\pagestyle{empty}
%\begin{document}
%\setlength{\unitlength}{1.0 mm}
\begin{figure}[tbh]
\begin{center}
%
\end{center}
\vspace*{5 mm}

\caption{}
\end{figure}
%\end{document}

\end{document}